\documentclass[twocolumn,twocolappendix]{aastex631} 

\usepackage{amsmath,amssymb,amsfonts,bm}
\usepackage{epstopdf}
\usepackage{epsfig}
\usepackage{natbib,caption2}
\usepackage{graphicx}   
\usepackage{float}
\usepackage{longtable}
\usepackage{graphics}
\usepackage{hyperref}
\usepackage{color}
\usepackage{calc}
\usepackage{threeparttable}

\newcommand \beq{\begin{equation}}
\newcommand \eeq{\end{equation}}
\newcommand \bey{\begin{eqnarray}}
\newcommand \eey{\end{eqnarray}}

\newcommand{\gsim}{\lower.5ex\hbox{$\; \buildrel > \over \sim \;$}}
\newcommand{\lsim}{\lower.5ex\hbox{$\; \buildrel < \over \sim \;$}}

\shortauthors{Wang et al.}

\begin{document}

\title{Disk Assembly of the Milky Way Suggested from the Time-resolved Chemical Abundance}
\email{ecwang16@ustc.edu.cn \\  jianhui.lian@ynu.edu.cn}

\author[0000-0003-1588-9394]{Enci Wang}
\affiliation{CAS Key Laboratory for Research in Galaxies and Cosmology, Department of Astronomy, University of Science and Technology of China, Hefei 230026, China}
\affiliation{School of Astronomy and Space Sciences, University of Science and Technology of China, Hefei, 230026, China}

\author[0000-0001-5258-1466]{Jianhui Lian}
\affiliation{South-Western Institute for Astronomy Research, Yunnan University, Kunming, Yunnan 650091, China}

\author{Yingjie Peng}
\affiliation{Department of Astronomy, School of Physics, Peking University, 5 Yiheyuan Road, Beijing 100871, China}
\affiliation{Kavli Institute for Astronomy and Astrophysics, Peking University, 5 Yiheyuan Road, Beijing 100871, China}

\author[0000-0002-9373-3865]{Xin Wang}
\affiliation{School of Astronomy and Space Science, University of Chinese Academy of Sciences (UCAS), Beijing 100049, China}
\affiliation{National Astronomical Observatories, Chinese Academy of Sciences, Beijing 100101, China}
\affiliation{Institute for Frontiers in Astronomy and Astrophysics, Beijing Normal University,  Beijing 102206, China}




\begin{abstract}

Both simulations and observations suggest that the disk assembly of galaxies is governed by the interplay between coplanar gas inflow, ex-planar gas outflow and in-situ star formation on the disk, known as the leaky accretion disk.  This scenario predicts a strong connection between radial distributions of star formation and chemical abundances. The Milky Way, being the sole galaxy where we can reliably measure star formation histories and the corresponding temporally-resolved chemical abundances with individual stars, provides a unique opportunity to scrutinize this  scenario.
Based on the recent large spectroscopic and photometric surveys of Milky Way stars, we obtain the radial profiles of magnesium abundance ([Mg/H]) and star formation rate (SFR) surface density at different lookback time. We find the radial profiles of [Mg/H] can be well-reproduced using the leaky accretion disk model with only two free parameters for stars formed within 4 Gyr, as well as the flattening at large radii of metallicity profiles traced by HII regions and Cepheids. 
Furthermore, the constraint effective yield of the Milky Way and nearby galaxies show broad consistency with the theoretical predictions from stellar chemical evolution model with a mass-loading factor of 0-2.  
These results support that the recent assembly of the Milky Way adheres to the leaky accretion disk scenario, bridging the disk formation of our home galaxy to the big picture of disk formation in the Universe.  
\end{abstract}


\section{Introduction}

While disk galaxies are widely seen in the local Universe and beyond, the mechanisms driving their growth and concurrent metal enrichment remain inadequately understood. 
Insight from simulations and observations is crucial in unraveling how disks accrete gas to sustain star formation. There is growing evidence in simulations that the inflowing gas is largely coplanar and corotating with the gas disk, particularly at low redshift \citep{Stewart-11, Stern-20, Trapp-22}, while outflowing gas driven by stellar winds and galactic feedback leaves the galaxies preferentially from two sides of disks \citep{Nelson-19, Peroux-20}. This aligns with observational indications from mapping the circumgalactic medium with Mg II absorption or emission \citep{Kacprzak-12, Diamond-Stanic-16, Bielby-17, Peroux-17, Schroetter-19, Guo-23}. 


Motivated by this idea, \cite{Wang-22a} built a disk formation model that simply treats the gas disk of star-forming galaxies as an idealized ``modified'' (or ``leaking") accretion disk (MAD).  
\cite{Wang-22a} explored the mechanisms underlying coplanar inflow in a gas disk, focusing on viscous processes. 
This study demonstrates that magnetic viscosity can naturally generate observed exponential profiles of star formation rate (SFR) surface density in galactic disks, as previously observed \citep[e.g.][]{Gonzalez-Delgado-16, Wang-23}.  
Based on MAD scenario, \cite{Wang-22b} investigated the metallicity profile in MAD scenario and found that the MAD model can explain the radial profiles of gas-phase metallicity for nearby galaxies. The negative radial gradient of metallicity is a natural consequence of the progressive enrichment of the gas by in-situ star formation as gas flows inwards through the disk. 

In the past decade, integral field spectroscopy surveys have mapped over ten thousand nearby galaxies \citep[e.g.][]{Croom-12, Bundy-15}, providing numerous spatially-resolved measurements, including gas-phase and stellar metallicity. However, measurements of the temporal evolution of metal abundance remain quite limited. For instance, oxygen abundance is commonly measured from the flux ratio of emission lines \citep[e.g.][]{Kewley-06}, which traces the instantaneous abundance for newly formed stars on a timescale of 10 Myr. Although spectral energy distribution fittings can, in principle, provide the temporal evolution of metallicity, the strong degeneracy between stellar age, metallicity, and dust attenuation renders the results unreliable \citep[e.g.][]{Conroy-13}.  The Milky Way, as a grand design disk galaxy, is a unique case to examine this disk formation scenario {\it across its lifetime} of disk formation and evolution. It is currently the only galaxy that can be  observationally resolved into individual stars, offering invaluable insights into the detailed constraints of disk formation and evolution.

Metallicity gradients of the Milky Way can be traced using various objects, including young and old stars, Cepheids, open clusters, HII regions, and planetary nebulae \citep{Luck-11, Balser-15, Magrini-17, Stanghellini-18, Anders-17, lian2023, Willett-23}. Specifically, \cite{Luck-11} obtained the iron and oxygen abundance of 101 Cepheids and derived the radial profiles of these metal abundances, tracing metallicity over a timescale of a few hundred Myr. \cite{Stanghellini-18} explored the time evolution of the metallicity gradient using planetary nebulae (PNe) with progenitors of different ages as metallicity probes, finding that the gradient from younger progenitor stars is steeper than that from older progenitor stars. \cite{Willett-23} investigated the metallicity gradient in the thin disc using 668 red-giant stars, finding a smooth evolution of the gradient from $-$0.07 dex/kpc in the youngest stars to $-$0.04 dex/kpc in stars older than 10 Gyr.

Models and simulations are widely used to understand the temporal evolution of metallicity gradients in the Milky Way and other extragalactic disk galaxies \citep[e.g.][]{Prantzos-00, Molla-05, Pilkington-12, Gibson-13, Grisoni-18, Vincenzo-18, Minchev-18, Belfiore-19, Wang-22b}. \cite{Chiappini-01} developed a chemical evolution model of the Milky Way, proposing two main formation episodes: the first forming the bulge over a short timescale, followed by a second forming the thin disk, with the timescale increasing as a function of Galactocentric radius \citep[also see][]{Matteucci-89, Pilkington-12, Grisoni-18}. They found that this ``inside-out" formation of the galactic disk, with a short halo formation timescale, offers the most plausible explanation for the Milky Way's disk. More recently, \cite{Grisoni-18} constructed a chemical evolution model of the Milky Way disk and suggested that the processes primarily influencing the formation of abundance gradients are the inside-out scenario, variable star formation efficiency, and radial gas flows. Assuming a negative radial gradient of gas-phase metallicity for most of the disk's lifetime, \cite{Minchev-18} used a semi-empirical approach to estimate Galactic birth radii for the Milky Way disk stars. They suggested that the gas-phase metallicity gradient flattens over time, from $-$0.15 dex/kpc at the beginning of disk formation to its measured present-day value of $-$0.07 dex/kpc. 

By examining hydrodynamical simulation models, \cite{Pilkington-12} found that the radial dependence of star formation efficiency over time drives variations in the metallicity gradients of simulated galaxies. They also discovered that systematic differences in sub-grid physics between various codes are responsible for setting these gradients. Similarly, \cite{Gibson-13} examined the role of energy feedback in shaping the distribution of metals and found that the strength of supernova feedback can significantly affect the metal distribution in galaxies. Using the Feedback in Realistic Environments \citep[FIRE;][]{Hopkins-14} simulations, \cite{Ma-17} investigated the metallicity gradients of high-redshift galaxies and found that strong negative metallicity gradients are only observed in galaxies with a rotating disk, while strongly perturbed galaxies with little rotation always have flat gradients.  


In this work, we revisit the temporal evolution of metal abundance in the Milky Way using the MAD model. Compared to previous models, the MAD model is very simple and produces the metallicity profiles of galaxies using only two free parameters (see Section \ref{sec:2.1}). 
In addition, APOGEE \citep[Apache Point Observatory Galactic Evolution Experiment;][]{majewski2017} survey releases robust and precise measurements of stellar parameters and elemental abundances for more than a half million stars across a full radial range from the Galactic center to the outer disk \citep{majewski2017, lian2023}. Therefore directly examining whether our Milky Way aligns with the MAD scenario is now possible.  

The paper is structured as follows. In Section \ref{sec:model}, we give a brief introduction of the MAD model and the data we used.  In Section \ref{sec:result}, we model the metallicity profiles of the Milky Way with MAD framework, and compare the Milky Way with MaNGA \citep[the Mapping Nearby Galaxies at APO;][]{Bundy-15} galaxies.  We summarize the results in Section \ref{sec:summary}.  Throughout this paper, the term ``metallicity'' specifically refers to the abundance of magnesium or oxygen ([Mg/H] or [O/H]), which represents the abundance of $\alpha$-elements. We assume a flat cold dark matter cosmology model with $\Omega_m=0.27$, $\Omega_\Lambda=0.73$ and $h=0.7$ when computing distance-dependent parameters.

\section{Method} \label{sec:model}

\subsection{The MAD model}  \label{sec:2.1}

Here we only give a brief description of the MAD model, and refer to the work of \cite{Wang-22a} and \cite{Wang-22b} for further details.  
In the MAD scenario, the radial inflow can be well constrained by star formation and the associated outflow with an assumption of disk quasi-equilibrium. This approach avoids arbitrary assumptions of the history of gas inflow, the radial gas distribution and the star formation law.
The inflow rate in equilibrium can be expressed as the integration of star formation and associated outflow rate: 
\begin{equation} \label{eq:1}
\Phi(r) =  \int_0^r 2\pi r' \cdot (1-R+\lambda)\Sigma_{\rm SFR}(r') dr' + \Phi(0)
\end{equation}
where $\Phi(r)$ is the net radial inflow rate, and $\Sigma_{\rm SFR}$ is the star formation rate surface density.  $R$ denotes the fraction of mass formed in new stars that is subsequently returned to the interstellar medium, and $\lambda$ is the ratio between mass outflowing rate and SFR, i.e. the mass-loading factor.  
The physical meaning of $\Phi(0)$ can be treated as the central mass sink onto the supermassive black hole and its associated outflow, which has negligible influence on the metallicity profiles \citep{Wang-22b}.  

In the MAD scenario the gas-phase metallicity $Z(r)$ can be written as: 
\begin{equation} \label{eq:2}
Z(r) = \int_{+\infty}^{r} -\frac{2\pi r' \cdot y\cdot \Sigma_{\rm SFR}(r')}{\Phi(r')} dr' + Z_0 
\end{equation}
where $Z_0$ is the original metallicity of inflowing gas,  and $y$ is the yield, i.e. the mass of metals returned to the interstellar medium per unit mass of formed stars. 
According to Equation \ref{eq:1} and \ref{eq:2}, if the $\Sigma_{\rm SFR}(r)$ is given, the $Z(r)$ profile can be determined with only two free parameters, $Z_0$ and effective yield, defined as \citep{Wang-21}:  
\begin{equation} \label{eq:20}
y_{\rm eff} = y/(1-R+\lambda). 
\end{equation}

Given that the lifetimes of massive stars that end as core-collapse supernovae (SNe) \citep[a few million years;][]{Roy-95} are negligible compared to the lifetime of galaxies, we adopt the instantaneous metal-enrichment and instantaneous mixing approximations. This approach effectively disregards the ejection of metals by Type-Ia SNe \citep{Mannucci-05}, a common assumption in many chemical evolution models. Following a single episode of star formation, over  90\% of the total released oxygen is expelled within 20 Myr by core-collapse SNe \citep[see figure 1 in][]{Maiolino-19}, while the contribution from Type-Ia SNe is minimal.

Therefore, the assumption of instantaneous metal-enrichment is reasonable for most of $\alpha$-elements but not for other elements such as carbon, nitrogen, and iron, because the production timescales for these elements are much longer (on the order of a Gyr). Additionally, the contribution to the production of these elements by Type-Ia SNe can be as significant as a few tens of percent \citep{Maiolino-19}.
In this work we use Magnesium, as a representative of $\alpha-$elements, rather Oxygen, because Magnesium abundance is the most reliable one among the $\alpha$-elements that we can obtain from the spectrum of stars. 

\subsection{Data of the Milky Way} \label{sec:sample}

\begin{figure*}
	\centering
	\includegraphics[width=8.8cm]{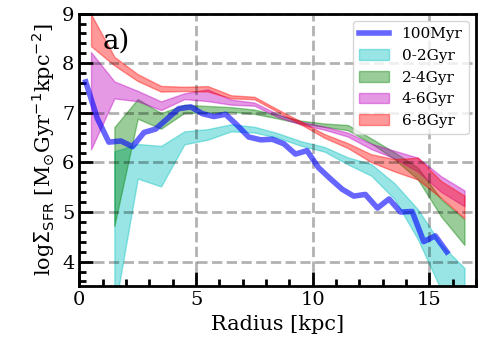}
    \includegraphics[width=8.8cm]{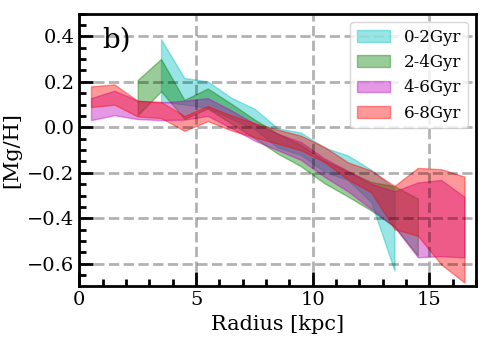}
	\caption{ Panel a): The $\Sigma_{\rm SFR}$ profile derived from individual stars of different ages from APOGEE survey.  The blue line shows the $\Sigma_{\rm SFR}$ profile estimated from the HiGAL observations \citep{Elia-22}.  Panel b): The corresponding [Mg/H] profiles for stars of different ages based on APOGEE survey. 
 } 
	\label{fig:SFR_prof}
\end{figure*}

The light-weighted average [Mg/H] profiles of stellar populations in different ages are derived utilizing APOGEE data from SDSS Data Release 17 \citep{abdurrouf2021}, following the same procedure to derive the [Fe/H] profile in \cite{lian2023}.  APOGEE is a near-infrared, high-resolution spectroscopic survey that has provided high-quality measurements of stellar parameters and elemental abundances (e.g., [Fe/H], [Mg/Fe]) for approximately 0.6 million stars in nearly 1,000 discrete fields.  Based on the density distribution of mono-abundance ([Mg/Fe] and [Fe/H]) populations from \cite{lian2022b}, we derive the integrated [Mg/H] profiles in two steps: we first integrate the 3D density distributions in the vertical direction and obtain the surface luminosity density profile for each mono-abundance population (MAP). These surface luminosity densities are then used as weight to calculate the average light-weighted [Mg/H] at each radii. For each MAP, we assume [Mg/H]$=$[Mg/Fe]$+$[Fe/H].  

The SFR surface density ($\Sigma_{\rm SFR}$) profiles of mono-age populations are estimated based on their living star mass surface density profile, which is derived by collapsing the 3D mass density distribution \citep{lian2022b} in the vertical direction and assuming no azimuthal variation.  The stellar ages and spectro-photometric distances have been derived by applying the astroNN deep-learning code to the spectroscopic data from APOGEE, asteroseismic data from APOKASC \citep{Pinsonneault-18, Mackereth-19}, and astrometric data from Gaia. 
These measurements are provided in the astroNN Value Added Catalog\footnote{https://data.sdss.org/sas/dr17/env/APOGEE\_ASTRO\_NN}, with typical age uncertainties of 30\% and distance uncertainties of 10\% \citep{Mackereth-19, Leung-19}.
Note that non-axisymmetric substructures, such as the bar and spiral arms, are not explicitly accounted for in the calculation of the surface density profile due to the size of the APOGEE sample. 
The mass fraction of living star to the total population formed is a function of stellar age and metallicity. We adopt the mass fractions calculated using stellar population models by \cite{maraston2005} and then apply them to the observed living star mass surface density to estimate the SFR surface density of mono-age populations.  
The radial profiles of [Mg/H] and $\Sigma_{\rm SFR}$ for stars of different ages are shown in Figure \ref{fig:SFR_prof}.  Similar profiles of the luminosity surface density of mono-age populations are presented in a parallel paper \citep{Lian-24}. 

In deriving the radial profiles of $\Sigma_{\rm SFR}$ and [Mg/H], no radial migration effect is assumed. However, this may not be fully realistic \citep{Schoenrich-Binney-09, Minchev-13, Kubryk-15}. 
For many years, it has been recognized that scattering by spiral structures and molecular clouds gradually heats the stellar disc, causing stars to move onto increasingly eccentric and inclined orbits. Specifically, \cite{Minchev-13} investigated the interplay between in situ chemical enrichment and radial migration and their impact on the radial gradients of [Fe/H]. They found significant flattening for older populations, while the metallicity gradient for the youngest stars (age $<$ 4 Gyr) is hardly affected at Galactocentric radii less than 12 kpc. Including the radial migration scheme in a chemical evolution model, \cite{Kubryk-15} quantitatively reproduced the main local properties of the thin and thick disk of the Milky Way: metallicity distributions and the ``two-branch'' behavior in the O/Fe vs. Fe/H relation. They found that radial migration nearly does not alter the radial distribution of the surface density of stars.


Simulations can, in principle, quantify the radial migration effect, although the resolution of stellar particles can influence stellar scattering and potentially enhance the radial migration effect \citep[e.g.][]{Elmegreen-13}. Using high-resolution hydrodynamical disk galaxy simulations, \cite{Sanchez-Blazquez-09} found that radial migration can increase the radial distribution of stars at large radii ($>$10 kpc) and flatten the metallicity gradients at a very weak level, less than 0.01 dex/kpc. \cite{Wang-23} investigated radial migration in TNG50 simulations \citep{Nelson-19b} and found that it can significantly flatten the age gradients of disk galaxies, which is clearly inconsistent with observations. Overall, we conclude that radial migration can potentially flatten the metallicity and age gradients, but it has a relatively weak effect on young stars, as expected. We will revisit this point in Section \ref{sec:3.1}.


In addition to these, we also extract the $\Sigma_{\rm SFR}$ and metallicity profiles of the Mikly Way at a more recent epoch from the literature. The $\Sigma_{\rm SFR}$ of a very recent epoch is taken from \cite{Elia-22}, which is constructed from the infrared clumps identified in the Hi-GAL \citep[the Herschel infrared Galactic Plane Survey;][]{Molinari-16} survey, shown in the blue line of Figure \ref{fig:SFR_prof}.  The $\Sigma_{\rm SFR}$ measurement reflects a timescale of 100 Myr \citep{Bell-03, Caplar-19}.  The [Mg/H] and [O/H] of this epoch are traced by the Cepheids and HII regions of the Milky Way, shown in Figure \ref{fig:z_prof_star}e and \ref{fig:z_prof_star}f respectively.  The metallicity of Cepheids represents a timescale of 100-400 Myr \citep{Pietrinferni-04, Pietrzynski-11}, and the metallicity of HII regions represents a timescale of 10 Myr \citep{Li-15, Caplar-19}.  
As both Magnesium and Oxygen are $\alpha$-elements, their radial abundance distribution closely aligns, for instance, indicated by Cepheids in Figure \ref{fig:z_prof_star}e. Consequently, we model in Section \ref{sec:result} the metallicity profile with combining the two elements together for Cepheids.  

There is a noticeable offset in metallicity profiles between HII regions and Cepheids, primarily attributed to different methods and calibrations used for measuring metallicity. We note that variations in methods and calibrations do result in significant discrepancies in metallicity measurements \citep{Kewley-06, Zhang-17, Easeman-24}. Therefore, comparing the metallicity profiles of different methods and of different data sets should be with caution. 

Throughout this work,  we assume the Oxygen abundance, 12+log(O/H), of the Sun is 8.69 \citep{Asplund-09}, and the solar metallicity is 0.02 \citep{Anders-89}. 

\subsection{MaNGA galaxies as reference} \label{sec:manga}

In comparing the Milky Way with nearby star-forming galaxies, we use a well-define sample of star-forming galaxies from \cite{Wang-19} as reference. This galaxy sample is originally selected from the SDSS (Sloan Digital Sky Survey) Data Release 14 \citep{Abolfathi-18}, with exclusions made for quenched galaxies, mergers, irregulars, and heavily disturbed galaxies. This careful selection results in a final MaNGA sample comprising 976 star-forming galaxies, providing a robust representation of normal main sequence galaxies. 

The gas-phase metallicity of MaNGA galaxies is computed with the {\tt N2S2H$\alpha$} diagnostic introduced by \cite{Dopita-16}.  The {\tt N2S2H$\alpha$} is expressed as ${\tt N2S2H\alpha} = \log([{\rm NII}]/[{\rm SII}]) + 0.264\log([{\rm NII}]/{\rm H}\alpha)$, where [NII] is the flux of [NII]$\lambda$6584 and [SII] is the total flux of [SII]$\lambda\lambda$6717,6731. 
The empirical relation of the oxygen abundance is then given by $12 + \log ({\rm O/H}) = 8.77 + {\tt N2S2H\alpha} + 0.45({\tt N2S2H\alpha}+ 0.3)^5$. 
The relevant emission lines, such as H$\alpha$, [NII]$\lambda$6584 and [SII]$\lambda\lambda$6717,6731, are located closely in wavelength, making the {\tt N2S2H$\alpha$} diagnostic insensitive to reddening.  Furthermore, \cite{Easeman-24} have examined various strong line diagnostics by comparing to sulphur $T_{\rm e}$-based metallicity measurements for a sample of HII regions in nearby galaxies. They found that {\tt N2S2H$\alpha$} diagnostic shows a near-linear relation with $T_{\rm e}$-based measurement, implying that the {\tt N2S2H$\alpha$} diagnostic is preferred when studying the distribution of metals within galaxies.

\section{Result} \label{sec:result}

\begin{figure*}
	\centering
	\includegraphics[width=10cm]{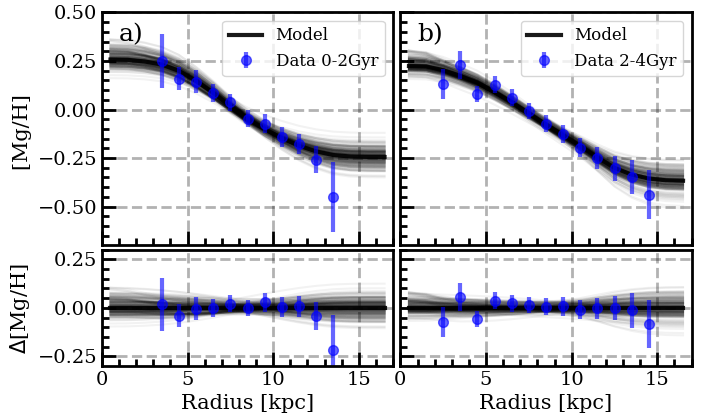}
    \includegraphics[width=6.07cm]{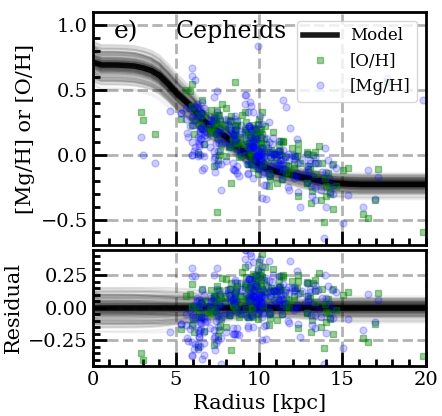}
    \includegraphics[width=10cm]{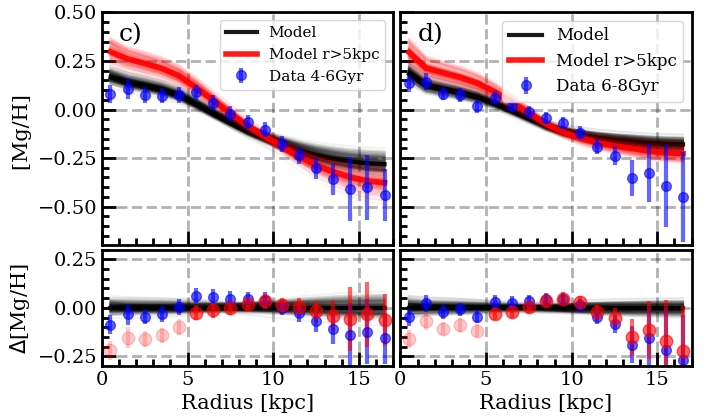}
    \includegraphics[width=6.07cm]{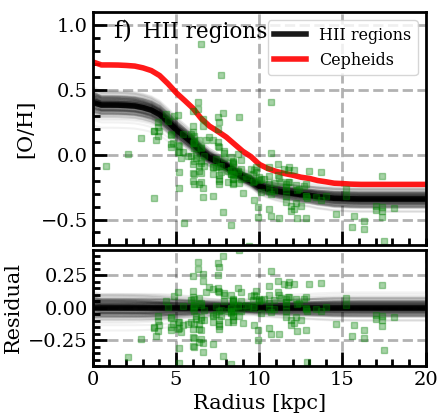}

	\caption{Panel a)-d): The radial profile of [Mg/H] for stars of different ages based on APOGEE survey \citep{lian2023}. The stars are divided into four age bins as in Figure \ref{fig:SFR_prof}, denoted in the legend of each panel.    Panel e)-f): Modeling the radial profile of [Mg/H] and/or [O/H] at a more recent timescale (from ten to a few hundred million years), as traced by Cepheids \citep{Luck-11, Genovali-15} and HII regions \citep{Balser-15}.   For comparison, in Panel f), we show the model of Cepheids in red line.  
    In each panel, the black thick line shows the best-fit model of MAD. The grey thin lines show the 200 models randomly chosen from MCMC chains, which represents the probability distribution of the preferred models.  Residuals with respect to the best-fit model are also displayed at the bottom. In panel c) and e), we also present the model with only fitting the data points at radii larger than 5 kpc, in red lines.  
 } 
	\label{fig:z_prof_star}
\end{figure*}

\begin{figure*}
	\centering
	\includegraphics[width=15cm]{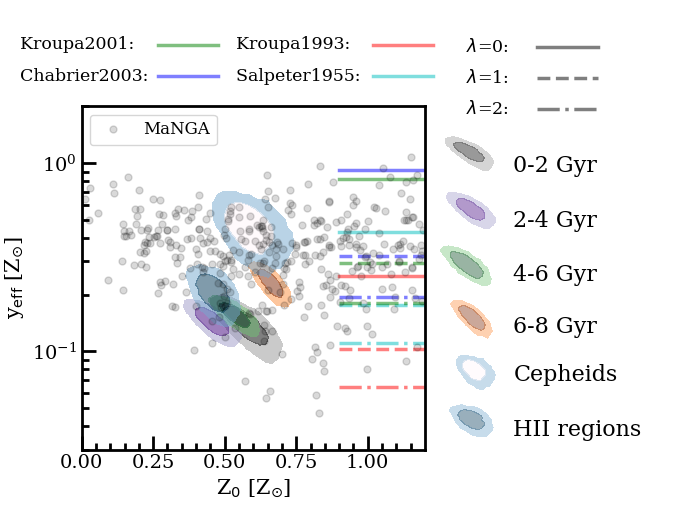}
	\caption{ The parameter spaces ($Z_0$ vs $y_{\rm eff}$) derived from the MCMC runs for metallicity profiles of different epochs (see Figure \ref{fig:z_prof_star}).  For each MCMC run, the contours show the 1-$\sigma$ region and 2-$\sigma$ region, respectively, of the probability distribution of $Z_0$ and $y_{\rm eff}$. 
    For comparison, we show the $Z_0$-$y_{\rm eff}$ distribution of nearby galaxies with $M_*>10^{10}M_{\odot}$ taken from the MaNGA survey, in grey dots.  In addition, we show the $y_{\rm eff}$ of Oxygen predicted from the chemical evolution models \citep{Vincenzo-16} with colored line segments, assuming various IMFs \citep{Salpeter-55, Kroupa-93, Kroupa-01, Chabrier-03} and mass-loading factors as denoted in the legend.   } 
	\label{fig:fit_para}
\end{figure*} 

\subsection{Modelling the Metallicity profile} \label{sec:3.1}

It is crucial to highlight that the Milky Way provides a unique perspective for examining models of cosmic disk formation.  In the case of extra-galactic galaxies, gas-phase metallicity and SFR are widely measured from the HII regions with a typical timescale of 10 million years. Stochastic variation may be significant at such short timescale, potentially challenging the assumption of quasi-equilibrium.  The long-term averaged quantities provided by the Milky Way are therefore valuable for understanding the disk assembly. In addition, the reconstructed $\Sigma_{\rm SFR}$ and metallicity profiles in Figure \ref{fig:SFR_prof} as a function of formation time of individual stars reveal the spatially resolved disk growth histories of the Milky Way.  This method is considered to be one of the most reliable ways to obtain galaxy star formation histories, avoiding complications like the degeneracy between age, metallicity, dust attenuation, and even the initial mass function (IMF) in common spectral fitting techniques.

By inputting the $\Sigma_{\rm SFR}$ profiles to Equation \ref{eq:1} and \ref{eq:2}, we employ Markov chain Monte Carlo (MCMC) method to obtain the preferred parameters with {\tt emcee} \citep{Foreman-Mackey-13} and examine whether this straightforward model can effectively match the observed metallicity profiles.    The Milky Way, like many low-redshift disk galaxies, exhibits negative metallicity gradients \citep{lian2023}. We do not consider the stars older than 8 Gyr, because the Milky Way might have been actively forming its bulge and thick disk rather than the thin disk at that time \citep{hasselquist2020}.  For stars formed within the last 0-2 and 2-4 Gyr, the MAD model predictions align almost perfectly with the [Mg/H] profiles of the Milky Way, as shown in Figure \ref{fig:z_prof_star}a and \ref{fig:z_prof_star}b.  This suggests that the Milky Way disk conforms to the MAD scenario.  In simpler terms, it implies that the assembly of the Milky Way is likely governed by the coplanar inflow in a quasi-equilibrium state over the last 4 Gyr. 

It is crucial not to dismiss this alignment as mere coincidence because the simplicity of the two-parameter model in Section \ref{sec:2.1} does not guarantee a good fit. Notably, when considering stars aged between 4-8 Gyr, the model struggles to capture the intricate details of the [Mg/H] profiles (see Figure \ref{fig:z_prof_star}c and \ref{fig:z_prof_star}d). This implies that the Milky Way might not have been in a state of quasi-equilibrium during that period. Evidence suggests a minor merger occurred approximately 6 Gyr ago \citep{ruiz-lara2020,lian2020a}, potentially destabilizing the gas disk and prompting gas flow (not enriched by star formation on the disk) toward the galactic center \citep{Li-08}. This is consistent with a central burst star formation around 4-8 Gyr ago, as depicted in Figure \ref{fig:SFR_prof}. 

Motivated by this, we specifically model the [Mg/H] profile of the Milky Way for stars of 4-8 Gyr, focusing solely on radii greater than 5 kpc.  This is shown in the red lines of Figure \ref{fig:z_prof_star}c and  \ref{fig:z_prof_star}d.  Interestingly, the [Mg/H] profile beyond 5kpc aligns well with the leaky accretion disk model for stars of 4-6 Gyr.  This alignment lends support to the proposed explanation mentioned earlier.  However, it is crucial to note that the [Mg/H] profile for stars of 6-8 Gyr still can not be well matched even only considering the radii greater than 5 kpc.  

For comparison, we show the [Mg/H] and/or [O/H] profiles of the Milky Way from a relatively recent period in Figure \ref{fig:z_prof_star}e and \ref{fig:z_prof_star}f, traced by Cepheids \citep{Luck-11, Genovali-15} and HII regions \citep{Balser-15}.  
By adopting the $\Sigma_{\rm SFR}$ estimated from infrared clumps, the fittings of MAD exhibit good agreement with the overall data points. Notably, the metallicity profiles of the Milky Way show a flattening at the larger radii, a feature not as evident in Figure \ref{fig:z_prof_star}a-d due to the limited range of radii but clearly seen in Figure \ref{fig:z_prof_star}e-f. This supports the notion that the Milky Way is presently constructing its disk within the MAD framework.

It is interesting to note that the [O/H] profile traced by HII regions is unaffected by the radial migration effect.  The good consistence between the metallicity profiles traced by HII regions and individual stars, shown in Figure \ref{fig:z_prof_star} and Figure \ref{fig:fit_para}, suggests that the effect of radial migration effect is weak for stars younger than 4 Gyr.

\subsection{Comparison with MaNGA galaxies} \label{sec:3.2}

\begin{figure*}
	\centering
	\includegraphics[width=8.8cm]{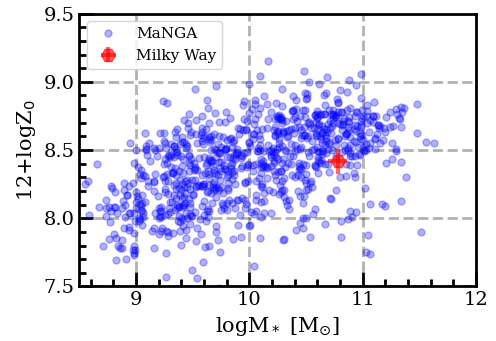}
        \includegraphics[width=8.8cm]{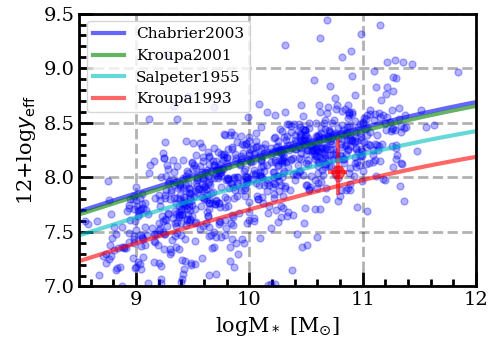}
	\caption{ The comparison of the Milky Way and MaNGA galaxies on the derived parameters, $Z_0$ and $y_{\rm eff}$.  In both panels, the Milky Way is shown in red hexagon. For the Milky Way, the value of each parameter is  the mean of the derived parameters from the six fittings (of different timescales) in Figure \ref{fig:fit_para}, with the error bar indicating the minimum ad maximum values.  In the right panel, the lines show the theoretical prediction of $y_{\rm eff}$ as a function of stellar mass with different IMFs, adopting the relation between mass-loading factor and stellar mass from \cite{Muratov-15}. } 
	\label{fig:com_manga}
\end{figure*} 

In a similar way, we perform fits to the metallicity profiles of MaNGA galaxies following the Equation \ref{eq:2}, incorporating the exponential form of $\Sigma_{\rm SFR}$ for simplicity \citep{Wang-22a}.  By inputting the observed scalelength of $\Sigma_{\rm SFR}$, we obtain the $Z_0$ and $y_{\rm eff}$ by directly fitting the metallicity profiles of MaNGA galaxies. The constrained parameter spaces for the metallicity profiles in Figure \ref{fig:z_prof_star} are shown in Figure \ref{fig:fit_para}.  For the Milky Way, the $Z_0$ is typically 0.5$Z_{\odot}$, and the $y_{\rm eff}$ is 0.2$Z_{\odot}$. Comparing with the MaNGA galaxies, the constrained parameters of the Milky Way fall within the range of typical values for disk galaxies. This suggests that the recent enrichment process in the Milky Way's disk is likely not unique among local galaxies.  The value of $Z_0$ is higher than the typical metallicity in cosmic-web filaments, suggesting that the inflowing gas is not directly from cosmic web, but likely the galaxy fountain enriched by gas recycling. 

Moreover, we show the $y_{\rm eff}$ values of Oxygen predicted from the chemical evolution models \citep{Vincenzo-16} in colored line segments in Figure \ref{fig:fit_para}, considering various assumptions of IMFs and mass-loading factors. It is worth emphasizing that although the values of $y_{\rm eff}$ are constrained by [Mg/H] rather than [O/H] in the this study, both elements primarily originate from core-collapse SNe, and the relative abundance to the Sun is expected to be similar (as illustrated in Figure \ref{fig:z_prof_star}e).  As shown in Figure \ref{fig:fit_para}, the fitted $y_{\rm eff}$ of the Milky Way and MaNGA galaxies are broadly in agreement with the $y_{\rm eff}$ obtained independently from stellar chemical evolution models with assuming mass-loading factor of 0-2 \citep{Muratov-15}.   This indicates that the values of fitted $y_{\rm eff}$ fall within a reasonable range. Additionally, if the value of yield is given, this method potentially provide a way to constrain mass-loading factor for galaxies.

Figure \ref{fig:com_manga} shows a comparison between the Milky Way and MaNGA galaxies in terms of the derived parameters, $Z_0$ and $y_{\rm eff}$, as a function of stellar mass. Overall, the Milky Way exhibits lower $Z_0$ and $y_{\rm eff}$ compared to MaNGA galaxies of similar mass, but it remains within the general distribution (i.e. within 2$\sigma$ region) in both panels of Figure \ref{fig:com_manga}.  In right panel, we show the theoretical prediction of $y_{\rm eff}$ as a function of stellar mass with different IMFs in colored lines.  To derive these lines, we adopt the relation between the mass-loading factor and stellar mass as $\lambda= 3.6(M_*/10^{10}M_{\odot})^{-0.35}$ \citep{Muratov-15}, based on FIRE simulations.  

It is noteworthy that the derived $y_{\rm eff}$-$M_*$ relation of MaNGA galaxies shows remarkable consistency, both in amplitude and slope, with the theoretical predictions of different IMFs.  This implies that our model provides an independent constraint on the value of mass-loading factor based on the metallicity profiles.  Even more intriguingly, the constrained $\lambda-M_*$ relation is broadly in agreement with the FIRE simulations \citep{Muratov-15}, reinforcing the validity and support for the idea presented in this work.

\section{Summary and Conclusion} \label{sec:summary}

Utilizing the recent spectroscopic and photometric surveys of Milky Way stars, we obtain the radial assembly histories and its metal enrichment histories of the thin disk.   This is probably the most reliable measurements of these two quantities for any individual disk galaxies, which provides an unique way to examine the MAD scenario as suggested from the observations and simulations.   Previous studies of metallicity profiles of the Milky Way usually focus on the galaxy itself with strong assumption of gas inflow, while we contextualize the Milky Way within the broader context of disk formation and derive the radial inflow rate from the observed $\Sigma_{\rm SFR}$, releasing the above strong assumption.   

We find that the radial profiles of [Mg/H] are well-reproduced under the MAD scenario with only two free parameters, for stars formed within the last 4 Gyr.  
Furthermore, this scenario successfully reproduces the observed metallicity profiles over a more recent timescale (from 10 Myr to a few hundred Myr) determined from the HII regions and Cepheids, and captures the distinct feature of profile flattening at larger radii. For stars of older ages, however, their [Mg/H] profiles show significant deviation from this scenario, suggesting that the earlier phases of formation involve more intricate processes, such as galaxy mergers.

Comparing with the MaNGA galaxies, the constrained parameters of the Milky Way fall within the range of typical values for disk galaxies, suggesting that the recent enrichment process in the thin disk of the Milky Way is likely not unique among local galaxies. In addition, the fitted $y_{\rm eff}$ of the Milky Way and MaNGA galaxies show broad agreement with the $y_{\rm eff}$ obtained independently from stellar chemical evolution models with a mass-loading factor of 0-2. This potentially provides a way to constrain the mass-loading factor of galaxies. 

The good agreement between the observational data and the modeled results strongly suggests that the assembly of the Milky Way has likely adhered to the leaky accretion disk scenario over the last 4 Gyr, where the co-planar inflow dominates the gas inflow and drives the disk growth. 
This study serves as a bridge, connecting the specifics of the Milky Way's disk formation to the broader understanding of disk formation in the Universe.

\section*{Acknowledgements}

E.W. acknowledges the support of Start-up Fund of the University of Science and Technology of China, Grant No. KY2030000200, and the Cyrus Chun Ying Tang Foundations. J.L. is supported by Yunnan Province Science and Technology Department under Grant Nos.
202105AE160021 and 202005AB16000 and the Start-up Fund of Yunnan University under Grant No. CY22623101. Y.P. acknowledges the support from the National Science Foundation of China (NSFC) grant Nos. 12125301, 12192220, 12192222, and the science research grants from the China Manned Space Project with No. CMS-CSST-2021-A07. X.W. is supported by the Fundamental Research Funds for the Central Universities, and the CAS Project for Young Scientists in Basic Research, Grant No. YSBR-062.

Funding for the Sloan Digital Sky Survey IV has been provided by the Alfred P. Sloan Foundation, the U.S. Department of Energy Office of Science, and the Participating Institutions. SDSS-IV acknowledges support and resources from the Center for High-Performance Computing at the University of Utah. The SDSS web site is www.sdss.org.

SDSS-IV is managed by the Astrophysical Research Consortium for the Participating Institutions of the SDSS Collaboration including the Brazilian Participation Group, the Carnegie Institution for Science,  Carnegie Mellon University, the Chilean Participation Group, the French Participation Group, Harvard-Smithsonian Center for Astrophysics,  Instituto de Astrof\'isica de Canarias, The Johns Hopkins University, Kavli Institute for the Physics and Mathematics of the Universe (IPMU) /  University of Tokyo, the Korean Participation Group, Lawrence Berkeley National Laboratory,  Leibniz Institut f\"ur Astrophysik Potsdam (AIP),  Max-Planck-Institut f\"ur Astronomie (MPIA Heidelberg), Max-Planck-Institut f\"ur Astrophysik (MPA Garching), Max-Planck-Institut f\"ur Extraterrestrische Physik (MPE), National Astronomical Observatories of China, New Mexico State University, New York University, University of Notre Dame, Observat\'ario Nacional / MCTI, The Ohio State University, Pennsylvania State University, Shanghai Astronomical Observatory, United Kingdom Participation Group, Universidad Nacional Aut\'onoma de M\'exico, University of Arizona, University of Colorado Boulder, University of Oxford, University of Portsmouth, University of Utah, University of Virginia, University of Washington, University of Wisconsin, Vanderbilt University, and Yale University.



\bibliographystyle{aasjournal}
\bibliography{reference}{}



\end{document}